\documentclass[prb,onecolumn,preprintnumbers,amsmath,amssymb,11pt]{revtex4-2}
\usepackage[utf8]{inputenc}
\usepackage{epsf}
\usepackage{graphicx}
\usepackage{latexsym}
\usepackage{amsmath}
\usepackage{amssymb}
\usepackage{amsfonts}
\usepackage{color}

\usepackage{enumitem}

\usepackage{xcolor}
\definecolor{darkblue}{rgb}{0.00, 0.00, 0.55}
\definecolor{darkmagenta}{rgb}{0.50, 0.00, 0.50}
\definecolor{darkcerulean}{rgb}{0.03, 0.27, 0.49}

\usepackage{amsmath}
\usepackage{amssymb}
\usepackage{graphicx,epsf}
\usepackage{hyperref}
\hypersetup{
    colorlinks         = true,
    citecolor          = darkblue,
    linkcolor          = darkblue,
    filecolor          = magenta,
    urlcolor           = darkmagenta,
    bookmarks             = false,
    unicode                  = true,
    bookmarksopen       = true,
    bookmarksopenlevel = 1
}

\begin{document}

\title{\Large{Visualization of Atomic Structures on Faceted and Nonflat Surfaces by the Difference-of-Gaussians Approach}}

\author{A. Yu. Aladyshkin$^{1-3, *}$, A. N. Chaika$^{4}$, V. N. Semenov$^{4}$, A. S. Aladyshkina$^{5}$, S. I. Bozhko$^{4}$,  A. M. Ionov$^{4}$}

\bigskip
\affiliation{$^1$Institute for Physics of Microstructures Russian Academy of Sciences, GSP-105, 603950 Nizhny Novgorod, Russia\\
$^2$Center for Advanced Mesoscience and Nanotechnology, Moscow Institute of Physics and Technology, Institutsky str. 9, 141700 Dolgoprudny, Russia\\
$^3$Lobachevsky State University of Nizhny Novgorod, Gagarin Av. 23, 603022 Nizhny Novgorod, Russia  \\
$^4$Osipyan Institute of Solid State Physics Russian Academy of Sciences, \mbox{Acad. Osipyan str. 2, 142432 Chernogolovka, Russia} \\
$^5$National Research University Higher School of Economics (HSE University), \mbox{Bolshaya Pecherskaya str. 25/12, Nizhny Novgorod, 603155, Russia} \\}


\maketitle



\vspace*{-5mm}

\section*{Abstract}

Detailed analysis of scanning probe microscopy (SPM) data acquired for faceted and non-flat surfaces is usually complicated due to the presence of a large number of surface areas tilted by large/variable angles relative to the scanning plane. As a consequence, standard methods of elimination of global or local slopes by either a plane subtraction or numerical differentiation seem to be ineffective. We demonstrate that a simple difference-of-Gaussians procedure provides output data corresponding to projection of a considered surface onto the scanning plane without undesirable contrast modifications. This method allows us to suppress small-scale noise, minimize effects of finite slopes in the SPM images along both fast and slow scanning directions and removes surface ripples without active participation of the operator. This method can be applied for fast on-the-fly visualization of experimental data, and for more detailed analysis, including high-precision determination of lattice parameters and angles between translation vectors for surface reconstruction of different terraces or surface domains. In order to estimate geometrical distortions introduced by our procedure, we compare the results obtained by the difference-of-Gaussians approach for the tilted surfaces with direct image rotation in 3D space.

$^*$ Corresponding author, e-mail address: aladyshkin@ipmras.ru

\section{Introduction}

Methods of scanning probe microscopy (SPM) are widely used in modern experimental physics and nanotechnology. In particular, scanning tunneling microscopy (STM) \cite{Binnig-82, Chen-93, Voigtlaender-15} and non-contact atomic-force microscopy (AFM) \cite{Ohnesorge-93, Voigtlaender-15, Voigtlaender-19, Morita-02} provide us valuable information concerning local atomic structure of surfaces. SPM methods are very useful for experimental investigations of low-dimensional and hybrid systems, as well as in studies of various surface reconstructions \cite{Oura-03}, that appear on single-crystal surfaces and involve one or several atomic layers. SPM data are often considered as the last argument in verification of various models of two-dimensional (2D) crystalline structures, in contrast to the results based on integral measurements such as low-energy electron diffraction (LEED) and reflected high-energy electron diffraction (RHEED). For example, the dimer-adatom-stacking fault model (so-called DAS-model) of Si(1\,1\,1)\mbox{$7\times 7$} reconstruction, introduced by Takayanagi \emph{et al.} \cite{Takayanagi-85} was generally accepted after high-resolution STM experiments by Binnig \emph{et al.} \cite{Binnig-83}.

Before the results of topography measurements acquired by SPM become suitable for structural analysis, they should be filtered from various artifacts keeping all valuable information undistorted. Typically, the sample surface is not ideally parallel to the scanning plane of the tip, and therefore raw topography images $z(x,y)$ obtained by the STM/AFM techniques have a global tilt. The tilt can be routinely excluded by means of plane subtraction,  with the plane defined by three reference points
\begin{eqnarray}
\label{Eq:def-alignment}
z'(x,y) = z(x,y) - (a\cdot x + b\cdot y + c),
\end{eqnarray}
where $a$, $b$ and $c$ are the plane parameters. Hereafter, we will refer to $z'(x,y)$ as the aligned topography image. The procedure of the plane subtraction (\ref{Eq:def-alignment}) is incorporated in commercial and freely distributed applications like WsXM \cite{WSXM}, Gwyddion \cite{Gwyddion} and others.

\begin{figure*}[t!]
\centering
\includegraphics[width=160 mm]{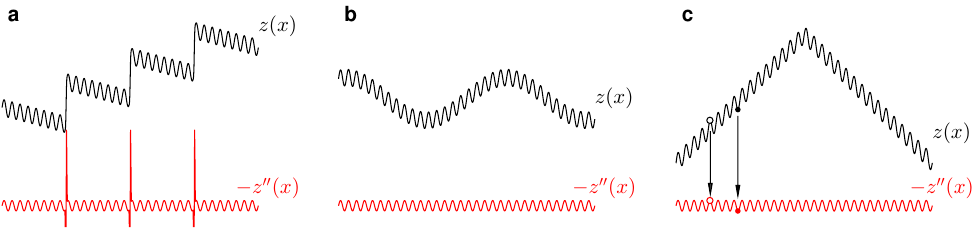}
\caption{Schematic presentation of one-dimensional topography images, corresponding to the vicinal surface with narrow terraces (panel a), non-flat surface (panel b) and 3D crystallite (panel c). The bottom row corresponds to the dependences $-z''(x)$ (red solid lines). Note that procedure (\ref{Main-idea-0}) corresponds to the projection of the considered surface onto the scanning plane. Additionally, this approach converts local maxima and minima on the raw signal $z(x)$ into local maxima and minima on the differential signal $-z''(x)$ (with experimental accuracy).
\label{Fig-01-artistic}}
\end{figure*}

\smallskip
However, procedure (\ref{Eq:def-alignment}) is badly conditioned and, as a result, ineffective in the following cases:

-- high-Miller-index (so-called vicinal) surfaces, consisting of narrow atomically flat terraces. Indeed, rather small width of the close-packed terraces (on the order of 5 nm) introduces substantial uncertainty in estimating the parameters of the plane (\ref{Eq:def-alignment}) and thus substantially complicates tilt removal. This problem as applied to stepped Si(5\,5\,6) and Si(5\,5\,7) surfaces was considered in a separate publication \cite{Aladyshkin-2024};

-- non-flat surfaces without any atomically flat terraces. This is a typical situation with rippled surfaces of graphene and graphene-like films. In this case one cannot eliminate pronounced large-scale curvature of the surface by the plane subtraction for any choice of the reference points;

-- faceted surfaces and three-dimensional (3D) crystallites. Since different faces have different individual tilt angles, the procedure (\ref{Eq:def-alignment}) cannot be applied for the entire topography image, but only for a specific face. Any preliminary analysis of the data and identification of interesting features on the faces during SPM experiments is hindered in this case;

-- unwanted image distortions caused by creep and nonlinear response of the piezo-scanner, temperature drift as well as by modification of the apex of the probe tip due to the migration of atoms in a strong electric field between the tip and the sample \cite{Tsong-1991}. These effects distort the topography image mainly along the slow-scanning direction leading to a nonlinear background.

\smallskip

In order to overcome the above issues one can apply the following strategies.

First, the procedure of direct numerical differentiation $\partial z/\partial x$ or $\partial z/\partial y$ makes it possible to exclude the global slope in a specific direction and analyze the periodic variations of the topography signal. For example, this approach was used for visualizing an atomic structure on the stepped Si(5\,5\,7) surface \cite{Kirakosian-APL-01,Chaika-JAP-09}. However, this approach apparently converts raw topography data into a picture with distorted contrast, hence it cannot be referred to as a pictorial presentation of a crystalline lattice (see the Supporting Information). In addition, numerical differentiation without additional filtering increases noise intensity, which could mask some details in the processed image.

Second, linear slopes in both directions and small-scale noise in raw topography image can be excluded by calculating the sum of the second-order partial derivatives from the filtered image $z^{\,}_{\rm filt}(x,y)$
\begin{eqnarray}
\label{Main-idea-0}
-\left(\frac{\partial^2}{\partial x^2} + \frac{\partial^2}{\partial y^2}\right)\, z^{\,}_{\rm filt}(x,y).
\end{eqnarray}
where the filtering of small-scale noise in the raw image can be done by any kind of smoothing, for example, the Gaussian blurring [see Eqs.~(\ref{Definition-gaussian})--(\ref{Definition-filt-f})]. This approach solves all mentioned difficulties (see figure~\ref{Fig-01-artistic}). We would like to note that procedure (\ref{Main-idea-0}) is useful for express analysis of the experimental data 'on-the-fly' and for deeper treatment of the STM data after completing the series of the SPM measurement.

\smallskip
Here we demonstrate that the approach (\ref{Main-idea-0}) can be realized using following procedures:

(i)  two-step procedure consisting of the Gaussian blurring of the raw image and direct numerical differentiation of the filtered image on a two-dimensional grid;

(ii) three-step procedure that starts with a calculation of the Fourier transform of the raw image, followed by its multiplication by the window function $|k|^2\cdot \exp(-|k|^2\sigma^2)$, and a calculation of the inverse Fourier transform (here $|k|=\sqrt{k_x^2+k_y^2}$ is the absolute value of the wave vector);

(iii) three-step procedure consisting of the difference between two filtered images prepared using the Gaussian blurring with different smoothing parameters [see Eq.~(\ref{Definition-DOG})]. We will refer to this procedure as the difference-of-Gaussians approach.

\smallskip
It should be noted that the methods involving calculations of the Laplacian of the analyzed image and the difference between two images smoothed by the Gaussian functions are well known and frequently used for digital image processing \cite{Gonsales-04,Kovasznay-53,Marr-80,Young-1987,Lindeberg-15,Assirati-2014}. These methods are typically applied in automatic detection of edges, interfaces and defects on half-tone images concerning, acquired by electron microscopy \cite{Krivanek-10,Voss-09,Misra-20} and magneto-resonant tomography \cite{Zhang-2016}. There are individual references of usage of the methods (i)--(iii) for the analysis of the SPM images \cite{Marsh-2018}, primarily focusing on the identification of edges/boundaries and the recognition of surface objects of a certain size. To the best of our knowledge, the difference-of-Gaussians approach has not been used for the visualization of crystalline lattices on tilted and curved surfaces including the determination of precise parameters of various surface reconstructions. The present study is aimed at introducing the difference-of-Gaussians approach to surface science community and demonstrate the advantages of applying this simple technique for surface reconstruction analysis in non-standard cases.

\smallskip
It is clear that the methods (i)--(iii) can be realized a step-by-step manner using several built-in functions in commercial and freely distributed applications dealing with visualization and analysis of the SPM data. As far as we are aware, the difference-of-Gaussians approach has not been incorporated yet in any modern applications in the form of a single convenient function. Alternatively, the users can decide to develop their own scenarios for data processing. For example, we use Matlab and Python programming languages to compose the scripts allowing us to remove the global slopes, and thus highlight the periodic component using the difference-of-Gaussians procedure in one click. In any case, this method substantially reduces processing time for preparing high-quality topography images in non-standard cases (from minutes to seconds) and does not require the active participation of the operator.

To illustrate the efficiency of the difference-of-Gaussians approach we consider atomically-resolved STM topography images:

(i) faceted surfaces of oxidized Cu single crystals \cite{Chaika-SurfSci-08} that contain tilted terraces at an angle on the order of $10-15^{\circ}$ with respect to the scanning plane;

(ii) non-flat surfaces of nanostructured graphene on SiC/Si(0\,0\,1) \cite{Chaika-NanoRes-2013, Chaika-Nanotech-2014, Chaika-Nano-2019}, with ripples (areas of a curved free-standing film) and domains with different lattice orientations;

(iii) non-flat surfaces of Au(1\,1\,1) films \cite{Muz-17} and Pb(1\,1\,1) films \cite{Aladyshkin-21,Aladyshkin-23} (see the Supporting Information).

\section{Model}

We begin with a discussion concerning the extraction of the periodic signals associated with 2D crystalline lattice from raw topography images $z(x,y)$. We assume that two-variable function $z(x,y)$ has periodic component, global slopes along both $x$- and $y$-axes, and small-scale noise, which impedes direct numerical differentiation.

It is easy to see that the convolution of an arbitrary function $z(x,y)$ with a two-dimensional Gaussian function
\begin{eqnarray}
\label{Definition-gaussian}
G^{\,}_{\sigma}(x,y) = \frac{1}{(\sqrt{2\pi}\sigma)^2}\,\exp\left(-\frac{(x^2+y^2)}{2\sigma^2}\right)
\end{eqnarray}
plays a role of a low-pass filter of spatial Fourier harmonics (here $\sigma$ is the standard deviation). Indeed, the Fourier transform of the filtered (or blurred) function
\begin{eqnarray}
\label{Definition-filt-f}
z^{\,}_{\rm filt}(x,y) = \iint z(x',y')\,G^{\,}_{\sigma}(x-x',y-y')\,dx'dy'
\end{eqnarray}
is the product of the Fourier transform of the analyzed function and the Fourier transform of the Gaussian function (\ref{Definition-gaussian})
\begin{multline}
\label{Definition-filt-fft}
\hat{z}^{\,}_{\rm filt}(k^{\,}_x,k^{\,}_y) =\mathcal{F}\big[z^{\,}_{\rm filt}(x,y)\big] = \frac{1}{2\pi} \iint z^{\,}_{\rm filt}(x,y)\,e^{-ik^{\,}_xx-ik^{\,}_yy}\,dxdy = \\ \hat{z}(k^{\,}_x,k^{\,}_y)\cdot\exp\left(-\frac{|k|^2 \sigma^2}{2}\right),
\end{multline}
where $|k|=\sqrt{k_x^2+k_y^2}$. Due to the factor $e^{-|k|^2\sigma^2/2}$, the spatial components with large $|k|\gtrsim \sigma^{-1}$ values are suppressed in the Fourier spectrum $\hat{z}(k^{\,}_x,k^{\,}_y)$ of the blurred function. The suppressed Fourier components apparently correspond to small-scale noise in the original function (with a typical period less than $\sigma$).

    \begin{figure*}[t!]
    \centering
    \includegraphics[width=160 mm]{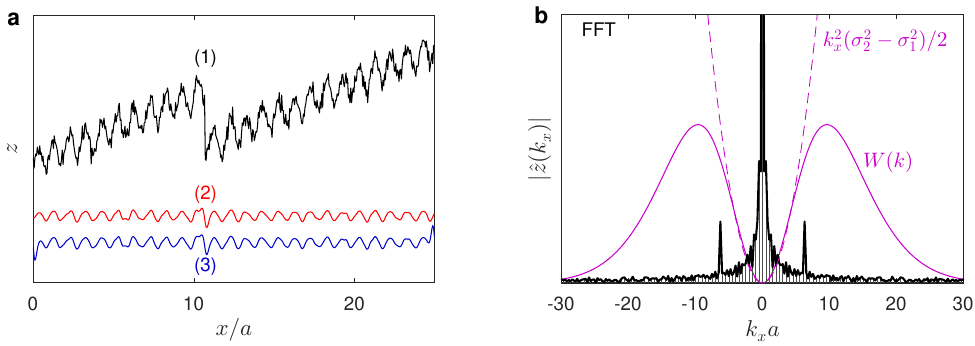}
    \caption{Illustrative presentation of the effective removal of the global slope and small-scale noise in one-dimensional case by filtering in the reciprocal space: \textbf{a}~-- Model noisy signal (1) and results of the filtering of periodic component by the difference-of-Gaussians approach in real space (2) and by the fast Fourier transforms (FFT) and band-pass filtering in the $k$-space (3) using the function $W(k)$ [see Eq.~(\ref{Definition-W})]. Lines 2 and 3 are shifted in the vertical direction for the sake of clarity. \textbf{b} -- Centered Fourier transform for the raw signal (thick solid line), the window function $W(k)$ (thin purple line), and the parabolic asymptote of $W(k)$ at small $|k|$ values according to Eq.~(\ref{Definition-DOG-asymptotes}) (dashed purple line). Lines 2 and 3 almost coincide in the inner region of the considered $x$-interval.
    \label{Fig-02-fft-differentiation}}
    \end{figure*}

\medskip
As mentioned in the Introduction, the global linear slope in a raw topography image in an arbitrary direction can be compensated by calculating the Laplacian (\emph{i.\,e.}, the sum of the second-order partial derivatives) of the blurred function
\begin{eqnarray}
\label{Definition-Laplacian}
-\Delta z^{\,}_{\rm filt}(x,y)=-\left(\frac{\partial^2 }{\partial x^2} + \frac{\partial^2 }{\partial y^2}\right)\,z^{\,}_{\rm filt}(x,y).
\end{eqnarray}
Since procedure (\ref{Definition-filt-f}) suppresses small-scale noise without affecting large-scale inhomogeneities such as linearly increasing components, procedure (\ref{Definition-Laplacian}) being applied to the blurred function $z^{\,}_{\rm filt}(x,y)$ allows us to highlight periodic components (figure~\ref{Fig-01-artistic}). Note that in typical SPM data the positions of local extrema in $z(x,y)$ and $-\Delta z^{\,}_{\rm filt}(x,y)$ for typical SPM data do coincide (within experimental accuracy). As a result, we can interpret Eq.~(\ref{Definition-Laplacian}) as the projection of the blurred surface onto the scanning $x-y$ plane (figure~\ref{Fig-01-artistic}c).

\medskip
Procedure (\ref{Definition-Laplacian}) of the direct numerical differentiation of the blurred function can be described in terms of the Fourier transform. Indeed, the Fourier transform of $-\Delta z^{\,}_{\rm filt}(x,y)$ is expressed via the Fourier transform of the raw function as follows
\begin{eqnarray}
\label{Definition-filt-fft-2}
\mathcal{F}\left[-\left(\frac{\partial^2 }{\partial x^2} + \frac{\partial^2 }{\partial y^2}\right)\,z^{\,}_{\rm filt}(x,y)\right] = \hat{z}(k^{\,}_x,k^{\,}_y)\cdot |k|^2 \cdot \exp\left(-\frac{|k|^2 \sigma^2}{2}\right).
\end{eqnarray}
In other words, one can get the similar result (\ref{Definition-Laplacian}) by taking the Fourier transform of the raw function $\hat{z}(k^{\,}_x,k^{\,}_y)$ and multiplying it by the window function
\begin{eqnarray}
\label{Definition-W}
W(k^{\,}_x, k^{\,}_y) = |k|^2\,\exp\left(-\frac{|k|^2\sigma^2}{2}\right),
\end{eqnarray}
which is effectively the same as applying a continuous band-pass filter in the $k$-domain (figure~\ref{Fig-02-fft-differentiation}b), and performing the inverse Fourier transform. The band-pass filter (\ref{Definition-W}) is not included in the list of available window functions for the Fourier transform in modern applications for SPM data analysis. Distortions caused by a piece-wise band-pass filtering in the $k$-domain are summarized in figures S1-S2 in the Supporting Information.

\medskip
We would like to show that the difference between two convolutions of a noisy function $z(x,y)$ with two Gaussian functions of different widths $\sigma^{\,}_1$ and $\sigma^{\,}_2>\sigma^{\,}_1$
\begin{multline}
\label{Definition-DOG}
D(x,y) \equiv \iint z(x',y')\,G^{\,}_{\sigma_1}(x-x',y-y')\,dx'dy' - \iint z(x',y')\,G^{\,}_{\sigma_2}(x-x',y-y')\,dx'dy'
\end{multline}
is proportional to the Laplacian of the blurred function (\ref{Definition-Laplacian}). According to Eq.~(\ref{Definition-filt-fft}), the Fourier transform of the differential signal $D(x,y)$ can be written in the form
\begin{eqnarray}
\nonumber
\hat{D}(k^{\,}_x,k^{\,}_y) = \hat{z}(k^{\,}_x,k^{\,}_y)\cdot W(k^{\,}_x,k^{\,}_y),
\end{eqnarray}
where
\begin{eqnarray}
\nonumber
W(k^{\,}_x,k^{\,}_y) = \exp\left(-\frac{|k|^2\sigma^2_1}{2}\right) - \exp\left(-\frac{|k|^2\sigma^2_2}{2}\right)
\end{eqnarray}
is the window function in the reciprocal space. Taylor series expansion of the exponents $e^{-|k|^2\sigma^2_1/2}$ and $e^{-|k|^2\sigma^2_2/2}$ results in the following asymptotic expressions
    \begin{eqnarray}
    \label{Definition-DOG-asymptotes}
    W(k^{\,}_x,k^{\,}_y) \simeq \left\{%
    \begin{array}{cc}
    |k|^2\,(\sigma^2_2-\sigma^2_1)/2 & \hbox{at $|k|\sigma^{\,}_{1,2}\ll 1$;} \\ [2mm]
    \exp\left(-|k|^2\sigma^2_1/2\right) & \hbox{at $|k|\sigma^{\,}_1\gtrsim 1$.} \\
    \end{array}%
    \right.
    \end{eqnarray}
We conclude that the difference-of-Gaussians procedure (\ref{Definition-DOG}) is equivalent to the suppression of small-scale noise with large $|k|$ values (since $W\propto e^{-|k|^2\sigma^2_1}$) accompanied by calculation of the Laplacian of large-scale inhomogeneities with small $|k|$ in the raw function $z(x,y)$ (since $W\propto |k|^2$).

Figure~\ref{Fig-02-fft-differentiation} illustrates the applicability of procedures (\ref{Definition-filt-fft-2}) and (\ref{Definition-DOG}) for extraction of the periodic component of a single-variable noisy function $z(x)$. The model trace, which mimics raw topography STM image containing sinusoidal corrugation, pronounced tilt, and small-scale noise, is shown in figure~\ref{Fig-02-fft-differentiation}a (line 1). The unfiltered Fourier transform of the raw signal trace $|\hat{z}(k^{\,}_x)|$ is presented in figure~\ref{Fig-02-fft-differentiation}b (thick black line). The main peak of $|\hat{z}(k^{\,}_x)|$ at $k^{\,}_x=0$ is associated with the finite slope of the raw function, while two symmetric maxima $k^{\,}_x=\pm 2\pi/a$ correspond to the contribution of the periodic component. The window function $W(k^{\,}_x)$ (thin purple line in figure~\ref{Fig-02-fft-differentiation}b) has a minimum at $k^{\,}_x=0$ and a maximum at intermediate $|k^{\,}_x|$ values. It demonstrates exponential decay for large $|k^{\,}_x|$ values. All these features of $W(k^{\,}_x)$ explain the suppression of both the finite slope and high-spatial-frequency noise. The inverse Fourier transform of $\hat{z}(k^{\,}_x)\cdot W(k^{\,}_x)$ is shown in figure ~\ref{Fig-02-fft-differentiation}a (line 3). The result of the direct calculations of the difference between two blurred functions according to Eq.~(\ref{Definition-DOG}) in real space (without using Fourier transforms) is shown in figure~\ref{Fig-02-fft-differentiation}a (line 2). As expected, the difference-of-Gaussians method introduces much smaller distortions near the edges of the considered interval (or in peripheral regions in two-dimensional images) in comparison with filtering of tilted images in the $k$-space. Thus the difference-of-Gaussians approach could be effective for the experimental investigations of 2D crystalline lattices on tilted and non-flat surfaces.

We would like to emphasize that the difference-of-Gaussians approach is rather robust and insensitive to a particular choice of smoothing parameters $\sigma^{\,}_1$ and  $\sigma^{\,}_2$, if we are interested in determination of periodicity and symmetry of surface corrugations in topography images. Surely, the amplitude of surface corrugations depends on the $\sigma^{\,}_2/\sigma^{\,}_1$ ratio, as well as the general appearance of the surface atomic features near mono- and multiatomic steps. The $\sigma^{\,}_1$ value should be smaller than the typical inter-atomic distance and it can be on the order of pixel size of the raw image. The $\sigma^{\,}_2$ value should be several times larger than $\sigma^{\,}_1$ but still less than the typical size of large-scale inhomogeneities (such as the width of terraces/faces, the period of surface modulation etc).

\section{Results and discussion}

All measurements were carried out in an ultrahigh vacuum (UHV) system, which includes a LAS-3000 (Riber) spectrometer equipped with Auger electron spectroscopy (AES) and low energy electron diffraction (LEED) techniques, and a GPI-300 scanning tunneling microscope ($\sum-$scan), operating at room temperature. The base pressure in the UHV chamber was below \mbox{$1\cdot 10^{-10}$\,mbar}. All topography images considered in this paper were acquired by means of STM in the regime of constant tunneling current $I$ and constant electrical potential of the sample $U$ with respect to the tip. As STM tips we used poly- and single crystalline tungsten wires etched in 2M of NaOH solution in water and then cleaned by successive electron and ion bombardment directly in the UHV chamber \cite{Chaika-SciRep-14}.

\subsection{Faceted Cu(1\,1\,5)--O surface}

\begin{figure}[h!]
\centering
\includegraphics[width=72 mm]{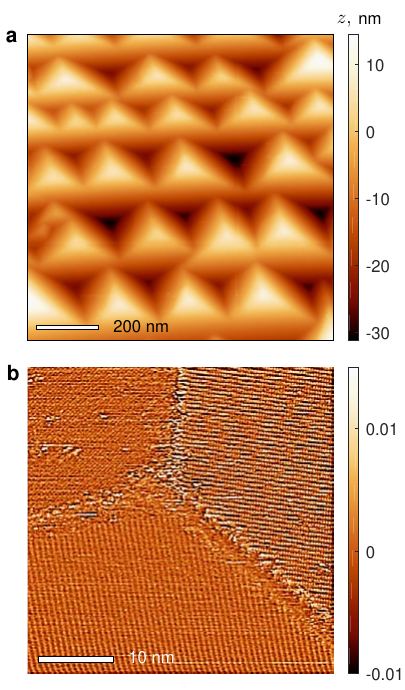}
\caption{\textbf{a}~-- Large-scale raw topography image of the \mbox{Cu(1\,1\,5)--O} surface, confirming the formation of the facetted surface (image size of \mbox{$1000\times 1000\,$nm$^2$}, tunneling voltage of \mbox{$U=-0.70\,$V}, mean tunneling current of \mbox{$I=150\,$pA}). \textbf{b}~-- Typical map of the differential signal $D(x,y)$ corresponding to the raw topography image of the \mbox{Cu(1\,1\,5)--O} surface acquired by means of procedure~(\ref{Definition-DOG}) without any other corrections (\mbox{$41\times 41\,$nm$^2$}, \mbox{$U=-10\,$mV}, \mbox{$I=120\,$pA}, the smoothing parameters are $\sigma^{\,}_1=0.08\,$nm and $\sigma^{\,}_2=0.16\,$nm (1 and 2 pixels, respectively).
\label{Fig03}}
\end{figure}

The procedure of preparation of oxygen-induced reconstructions on the Cu(1\,1\,5) surface was described in \cite{Chaika-SurfSci-08} in detail. In brief, copper single crystals with the (1\,1\,5) surface termination were polished and then annealed in an oxygen atmosphere ($T\simeq 1050^{\circ}$C and $p \simeq 6.5\cdot 10^{-4}\,$mbar) for 30 hours to remove organic and sulfur contaminants. After loading the sample into the UHV STM chamber, we use a series of consecutive Ar$^+$ bombardment ($p\simeq 5\cdot 10^{-5}\,$mbar, $E=550\,$eV) and annealing at moderate temperatures ($T\simeq 300^{\circ}C$ and \mbox{$p\simeq1\cdot 10^{-10}$\,mbar}) to get clean stepped Cu(1\,1\,5) surface. Faceting of the Cu(1\,1\,5) surface and an emergence of oxygen-induced reconstructions take place at temperatures close to $T\simeq 600^{\circ}$C due to oxygen diffusing out from bulk to the surface. According to the AES data, the reconstructed \mbox{Cu(1\,1\,5)--O} surface has 0.5 monolayer of oxygen \cite{Chaika-SurfSci-08}. A typical large-scale topography image of the \mbox{Cu(1\,1\,5)--O} surface is shown in figure~\ref{Fig03}a. According to our STM data and the results presented in Refs. \cite{Reiter-SurfSci-1996,Taglauer-Nucl-1996,Reinecke-SurfSci-2000, Reinecke-02,Walko-SurfRevLett-1999,Walko-01,Sotto-SurfSci-1992}, the presence of oxygen on the Cu(1\,1\,5) surface results in formation of two faces of the \{0\,1\,4\} family, as well as \{1\,1\,3\} and \{0\,0\,1\} faces (see figure S3 in the Supporting Information). The high-Miller-index faces of Cu crystals are often considered as perspective surfaces in catalysis studies and surface-specific reactivity \cite{Chang,Osada,Dhakar}.

Figure~\ref{Fig03}b shows a typical map of the differential signal corresponding to small-scale raw topography image of the reconstructed Cu(1\,1\,5)--O surface. Since the procedure described by Eq.~(\ref{Definition-DOG}) effectively removes the global tilts along both directions, the map of the differential signal $D(x,y)$ should correspond to the projection of the atomic lattices on the facets on the scanning $x-y$ plane. Apparently, the difference-of-Gaussians approach and the standard method of the plane subtraction being applied on a particular face of the faceted surface will give the same result. We emphasized that the difference-of-Gaussians approach helps to visualize easily the crystalline lattices at all faces of 3D nanocrystals \emph{simultaneously} provided that the crystalline lattices are recognizable. For example, one can simultaneous visualize the crystalline lattices on three different faces of the faceted Cu(1\,1\,5)-O surface (figure~\ref{Fig03}b).

\begin{figure}[h!]
\centering
\includegraphics[width=75 mm]{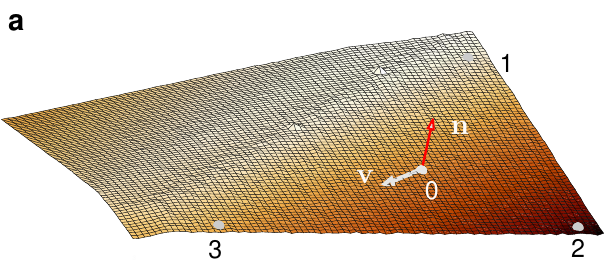}
\includegraphics[width=57 mm]{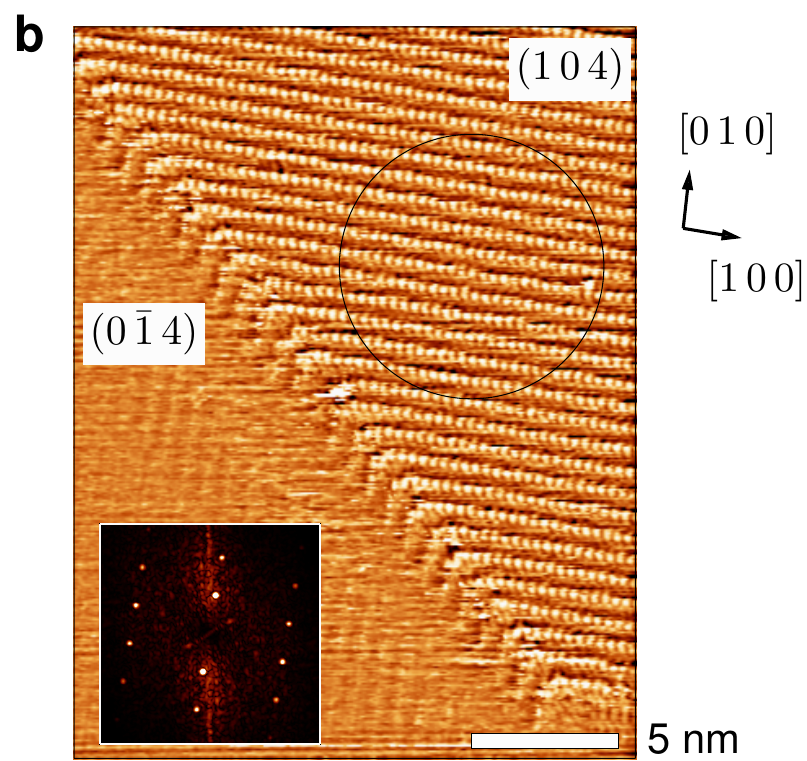}
\caption{\textbf{a} -- 3D presentation of the raw topography image of the \mbox{Cu(1\,1\,5)--O} surface ($19.4\times 24.9$\,nm$^2$, $U=-10\,$mV, $I=150\,$pA). Vectors ${\bf n}$ and ${\bf v}$ are the unit normal vector and the unit tangent vector along the level curve $z(x,y)=z^{\,}_0$, respectively. Both vectors are built at the middle point $(x^{\,}_0,y^{\,}_0,z^{\,}_0)$. \mbox{\textbf{b} --} Map of the differential signal $D(x,y)$ corresponding to the image in the panel a and acquired by means of Eq.~(\ref{Definition-DOG}) without any other corrections ($\sigma^{\,}_1=0.16\,$nm and $\sigma^{\,}_2=4\sigma^{\,}_1$). The area inside the circle is used for Fourier analysis (see the inset and figure~\ref{Fig-06-FFT}). Both images were affine transformed to restore the orthogonality of axes.
\label{Fig04}}
\end{figure}

As mentioned in the Introduction, the linear slope in a topography image can be removed by numerical calculation of the first derivative ($\partial z/\partial x$ or $\partial z/\partial y$). Although this method does not change the periodicity, it affects the spatial distribution of the contrast for a processed image. As a result, local maxima and minima in the $\partial z/\partial x$ map do not coincide with the positions of the atomic features in the raw image (see figures S4-S6 in the Supporting Information). This complicates further detailed analysis of experimental SPM data.

Figure~\ref{Fig04}a shows a small-scale topography image of the reconstructed Cu(1\,1\,5)--O surface in 3D geometry. One can easily see two faces, oriented at angles of $6^{\circ}$  and $14^{\circ}$ (left and right areas in this image) with respect to the scanning plane. The dihedral angle between these facets is equal to $20.5^{\circ}$. This value is close to the theoretical estimate of $19.7^{\circ}$ for the angle between the planes characterized by the Miller indices  $(1\,0\,4)$ and $(0\,\bar{1}\,4)$. The map of the differential signal $D(x,y)$ for the topography image is depicted in figure~\ref{Fig04}b. This figure clearly shows two series of perpendicular rows corresponding to the $c(4\times 1)$ reconstruction on both faces.

\begin{figure}[h!]
\centering
\includegraphics[width=62 mm]{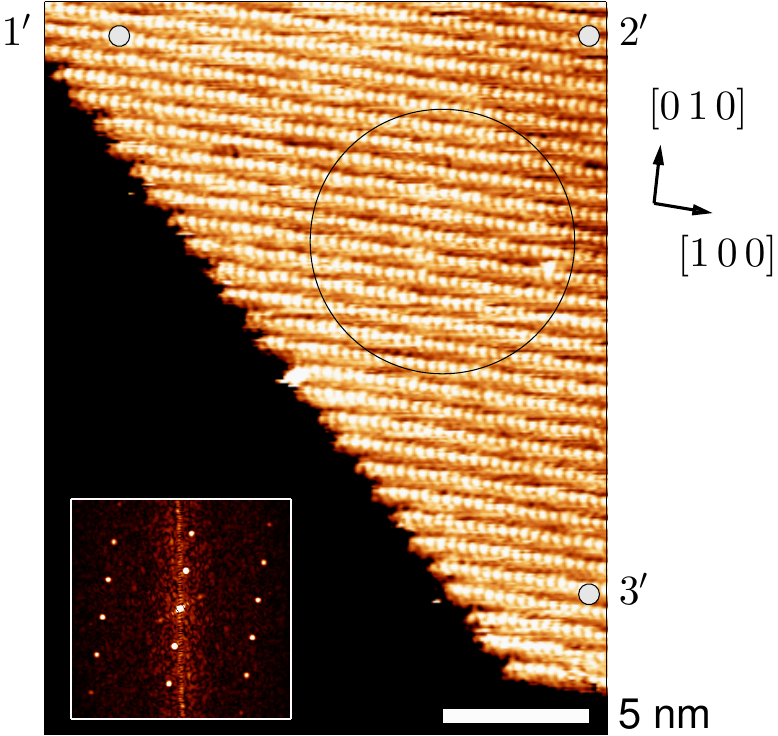}
\caption{Rotated topography image $z'(x',y')$ obtained from the raw topography image $z(x,y)$ shown in figure~\ref{Fig04}a after the rotation at an angle of $\gamma=14.7^{\circ}$ around the vector ${\bf v}$ by means of procedure (\ref{Eq:rotation-1})--(\ref{Eq:rotation-2}). Here, $1'$, $2'$, and $3'$ are the reference points $1, 2$ and $3$ after the rotation. The area inside the circle is used for the Fourier analysis (see the inset and figure~\ref{Fig-06-FFT}).
\label{Fig-05-rotated-image}}
\end{figure}

It is clear that the lattice parameters for a 2D atomic structure in the direction along the faceted surface and for the projection of the atomic structure on the $x-y$ plane should differ: the larger the slope of the given facet with respect to the scanning plane, the larger the difference. In order to confirm this conclusion, we compare the map of the differential signal  $D(x,y)$ (figure~\ref{Fig04}b) and the topography image $z'(x',y')$ (figure~\ref{Fig-05-rotated-image}), which can be obtained by rotation of the raw image $z(x,y)$ in 3D space. Let us introduce the reference points  $(x^{\,}_1,y^{\,}_1,z^{\,}_1)$,  $(x^{\,}_2,y^{\,}_2,z^{\,}_2)$, and $(x^{\,}_3,y^{\,}_3,z^{\,}_3)$, which determine the plane for the given face (figure~\ref{Fig04}a); $(x^{\,}_0,y^{\,}_0,z^{\,}_0)$ is the position of the middle point of the triangle $1-2-3$. Let ${\bf n}$ be the unit vector of normal for the plane $1-2-3$. We consider the level curve $z(x,y)=z^{\,}_0$, running through the point $(x^{\,}_0,y^{\,}_0,z^{\,}_0)$, and the unit tangent vector ${\boldsymbol v}= \big(v^{\,}_x, v^{\,}_y, v^{\,}_z\big)$ at this point (figure~\ref{Fig04}a). We will rotate the surface $z(x,y)$ by the angle $\gamma={\rm arccos}\,n^{\,}_z$ around the vector ${\bf v}$.

It is well known that the coordinates $(x',y',z')$ of each point of the rotated image can be expressed via the coordinates $(x,y,z)$ of the raw image as follows
\begin{eqnarray}
\label{Eq:rotation-1}
\left(
  \begin{array}{c}
    x' \\ [1mm]
    y' \\ [1mm]
    z' \\ [1mm]
  \end{array}
\right) = \hat{M} \left(
  \begin{array}{c}
    x \\ [1mm]
    y \\ [1mm]
    z \\ [1mm]
  \end{array}
\right),
\end{eqnarray}
where the rotation matrix $\hat{M}$ has the form
\begin{eqnarray}
\label{Eq:rotation-2}
\hat{M} = \left(
  \begin{array}{ccc}
    \cos\gamma + v_x^2\,(1-\cos\gamma) & v^{\,}_xv^{\,}_y\,(1-\cos\gamma) - v^{\,}_z\,\sin\gamma & v^{\,}_xv^{\,}_z\,(1-\cos\gamma) + v^{\,}_y\,\sin\gamma \\ [1mm]
    v^{\,}_xv^{\,}_y\,(1-\cos\gamma) + v^{\,}_z\,\sin\gamma & \cos\gamma + (1-\cos\gamma)\,v_y^2 & v^{\,}_yv^{\,}_z\,(1-\cos\gamma) - v^{\,}_x\,\sin\gamma \\ [1mm]
    v^{\,}_xv^{\,}_z\,(1-\cos\gamma) - v^{\,}_y\,\sin\gamma &  v^{\,}_yv^{\,}_z\,(1-\cos\gamma) + v^{\,}_x\,\sin\gamma & \cos\gamma + (1-\cos\gamma)\,v_z^2 \\ [1mm]
  \end{array}
\right).
\end{eqnarray}
An example of the rotation of the raw topography image is displayed in figure~\ref{Fig-05-rotated-image}.

\begin{figure}[h!]
\centering
\includegraphics[width=60 mm]{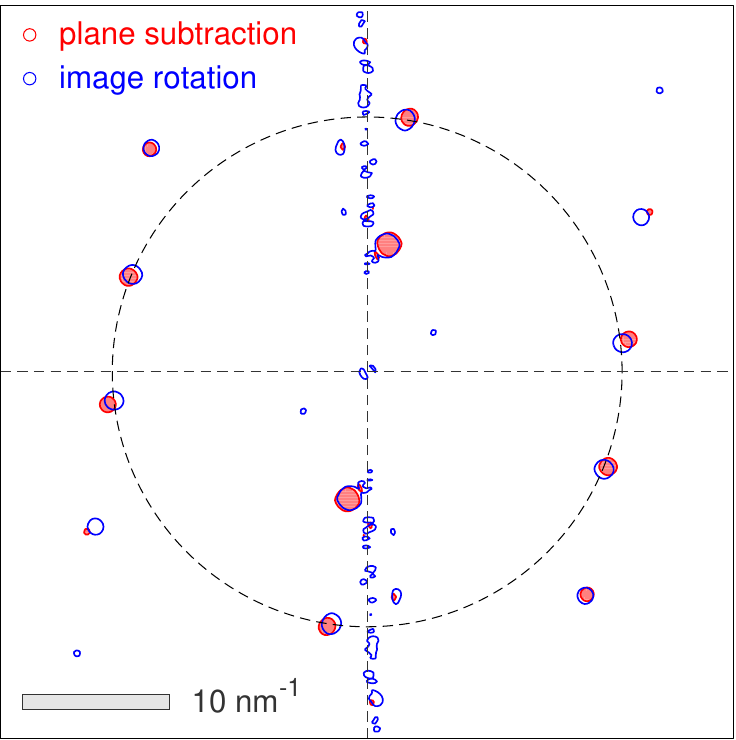}
\caption{Comparison of two Fourier transforms for the surface $c(4\times 1)$ reconstruction: filled level lines correspond to the projected topography image obtained by the difference-of-Gaussians approach (the inset in figure~\ref{Fig04}b); open level lines correspond to the rotated topography image (the inset in figure~\ref{Fig-05-rotated-image}).  The radius of the circle, that characterizes the $1\times 1$ lattice, is equal to  $k^{\,}_0=2\pi/a = 17.38\,$nm$^{-1}$, where $a = 0.361\,$nm is the lattice constant for bulk Cu.
\label{Fig-06-FFT}}
\end{figure}

The reciprocal space structures obtained by the fast Fourier transform for (i) the topography images aligned by plane subtraction, and (ii) image rotation are presented in figure~\ref{Fig-06-FFT}. The Fourier analysis shows the formation of a two-dimensional $c(4\times 1)$ lattice with a centered rectangular unit cell in both cases. Comparing mutual positions of the level lines, one can conclude that the finite slope of the faces results in (i) a noticeable shift of all Fourier peaks to larger $|k|$ values (at about 3-4\% for the face tilted by 15$^{\circ}$), and (ii) possible distortion of the symmetry of the 2D crystalline lattice. In other words, the geometrical distortions are on the order of $\cos^{-1}\alpha$, where $\alpha$ is the angle between the face and the scanning plane in radians. This means that the distortions introduced by the difference-of-Gaussians approach seem to be less than the natural width of the first-order Fourier maxima for faces tilted by $10^{\circ}$ or smaller.

\begin{figure}[h!]
\centering
\includegraphics[width=85 mm]{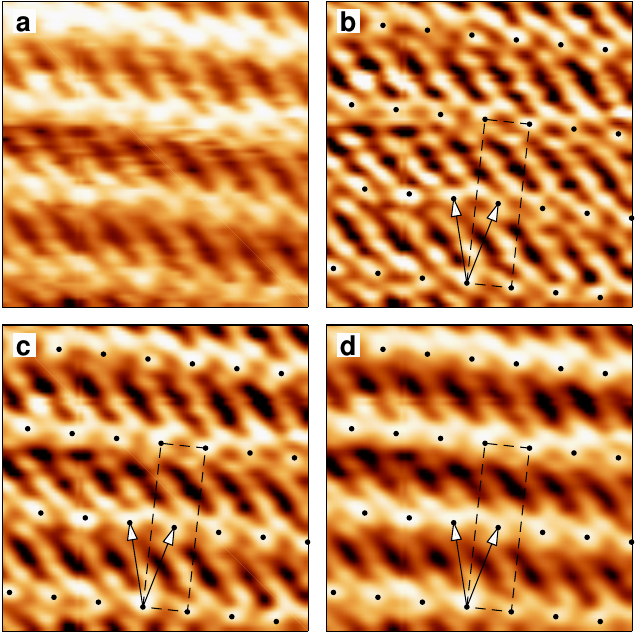}
\caption{\textbf{a} – Atomically resolved topography image of the \mbox{Cu(1\,0\,4)--O} face (image size of $2.5\times 2.5$\,nm$^2$, $U=-10\,$mV, $I=0.2\,$nA). \textbf{b -- d}~-- Maps of the differential signal $D(x,y)$ acquired according to Eq.~(\ref{Definition-DOG}), the smoothing parameters are $\sigma^{\,}_1=0.03\,$nm or 1\,pixel and $\sigma^{\,}_2=0.06\,$nm or 2 pixels (panel b), $\sigma^{\,}_2=0.13\,$nm or 4 pixels  (panel c), and $\sigma^{\,}_2=0.26\,$nm or 8 pixels (panel d). Two vectors correspond to the basic translation vectors for the $c(4\times 1)$ reconstruction; black dots show the expected positions of the atoms for this lattice; a dashed rectangle depicts the unit cell.
\label{Fig-07-diff-widths}}
\end{figure}

High-resolution raw topography image of the Cu(1\,0\,4)--O reconstruction is shown in figure~\ref{Fig-07-diff-widths}a. The pronounced gradient of the contrast in the vertical direction indicates the finite tilt of the sample. The maps of the differential signal $D(x,y)$ obtained by means of the difference-of-Gaussian procedure (\ref{Definition-DOG}) with different smoothing parameters $\sigma^{\,}_1$ and $\sigma^{\,}_2$ (panels b--d in figure~\ref{Fig-07-diff-widths}) demonstrate that this approach is quite effective for highlighting the atomic features resolved in STM experiments even for non-linear background (see also figures \ref{Fig-09-Graphene}--\ref{Fig-11-Graphene2} and figures S4--S8 in the Supporting Information). Indeed, in figure~\ref{Fig-07-diff-widths} one can easily see close-packed atomic rows and monatomic steps typical for the $c(4\times 1)$ lattice (see figure S3 in the Supporting Information). By changing the ratio $\sigma^{\,}_2/\sigma^{\,}_1$ we can enhance/diminish various details in the processed topography image. We conclude that the use of $\sigma^{\,}_2$ to $\sigma^{\,}_1$ ratio close to 2:1 can be especially useful for the analysis of stepped surfaces since it produces comparable contrast on all terrace rows and in the regions of multiple steps if they are present in the scanned surface area. This results in a simultaneous visualization of upper-lying step atomic rows and lower-lying down-step rows which cannot be achieved by plane subtraction. The comparison of the $D(x,y)$ maps obtained at different $\sigma^{\,}_2$ parameters proves that the difference-of-Gaussians approach is robust to the choice of $\sigma^{\,}_1$ and $\sigma^{\,}_2$ and provides correct lateral positions of the experimentally measured features in a wide range of $\sigma^{\,}_1$ and $\sigma^{\,}_2$.

\subsection{Nanostructured graphene on SiC/Si(001) surface}

\begin{figure}[h!]
\centering
\includegraphics[width=80 mm]{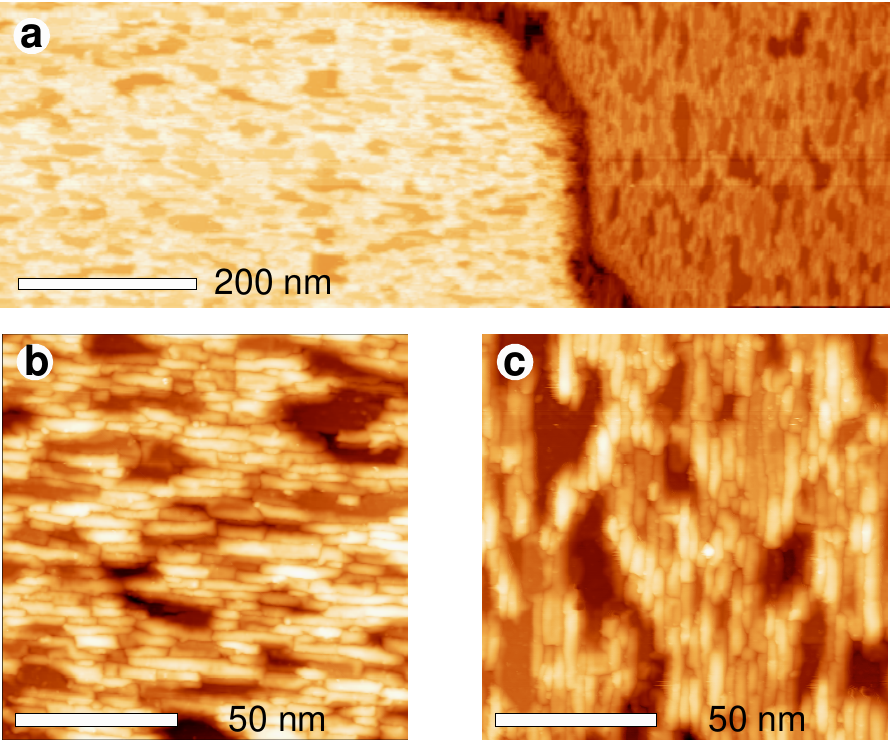}
\caption{\textbf{a} -- Overview topography image of a non-flat graphene film on the SiC/Si(001) surface, demonstrating two antiphase domains (image size of $1000\times 125$\,nm$^2$, $U=-2.1\,$V, $I=40\,$pA). \textbf{b, c}  -- Raw topography images of two areas on the non-flat graphene film with horizontally oriented (panel b) and vertically oriented (panel c) nanoribbons (image size of $125\times 125$\,nm$^2$, $U=-2\,$V, $I=70\,$pA (b) and $U=-1.4\,$V, $I=50\,$pA (c)).
\label{Fig-08-Graphene-overview}}
\end{figure}

Ultrathin layers of nanostructured graphene on \mbox{$\beta$-SiC/Si(001)} wafers were fabricated by means of high-temperature annealing in a UHV chamber (details of synthesis can be found in Refs. \cite{Chaika-NanoRes-2013,Chaika-Nanotech-2014}). It is well known that $\beta$-SiC thin films grown on the Si(001) single crystals consist of micron-size antiphase domains with bulk SiC square lattices rotated by $90^{\circ}$ relative to one another. As a result, mono- and multiatomic graphene layers appearing on top of the $\beta$-SiC films consisting of nanoribbons with lattice vectors rotated by $90^{\circ}$ and boundaries oriented predominantly along two perpendicular \{110\} directions \cite{Chaika-NanoRes-2013,Chaika-Nanotech-2014,Chaika-Nano-2019}. A typical large-scale STM image demonstrating two antiphase domains and graphene nanoribbons elongated in horizontal and vertical directions is shown in figure~\ref{Fig-08-Graphene-overview}a, while graphene nanoribbons within certain antiphase domains are shown in figures~\ref{Fig-08-Graphene-overview}b,c. Correct determination of the graphene lattice orientations in different domains and the atomic structure of the nanodomain boundaries, which can affect the electronic properties of the graphene overlayer, appears to be a non-trivial problem. Indeed, random deformation of the graphene layer caused by rather weak interaction with the \mbox{$\beta$-SiC/Si(001)} substrate (also known as atomic-scale rippling \cite{Fasolino}),  as well as the presence of domains with different orientation of the lattices, atomic defects in surface and subsurface layers leading to an additional modulation of the electronic states substantially complicate the problem of image processing. We would like to demonstrate the effectiveness of the difference-of-Gaussians approach in resolving this problem.

Figure~\ref{Fig-08-Graphene-overview}a shows the raw topography image of the nanostructured graphene layer on the \mbox{$\beta$-SiC/Si(001)} surface. The estimated standard deviation of the non-flat surface from an ideal plane (0.14 nm) is rather large and comparable with the distance between neighboring atoms in the graphene lattice (0.142\,nm). The map of the differential signal allows us to visualize small-scale variations of the intensity for all areas regardless of their absolute height and local slope (figure~\ref{Fig-08-Graphene-overview}b). The removal of the surface curvature reduces the standard deviation by the order of magnitude (down to 0.02\,nm): it becomes close to the natural level of corrugation for the flat areas of graphene or the flat surface of highly oriented pyrolytic graphite (HOPG) measured with a sharp STM probe.

\begin{figure}[h!]
\centering
\includegraphics[width=85 mm]{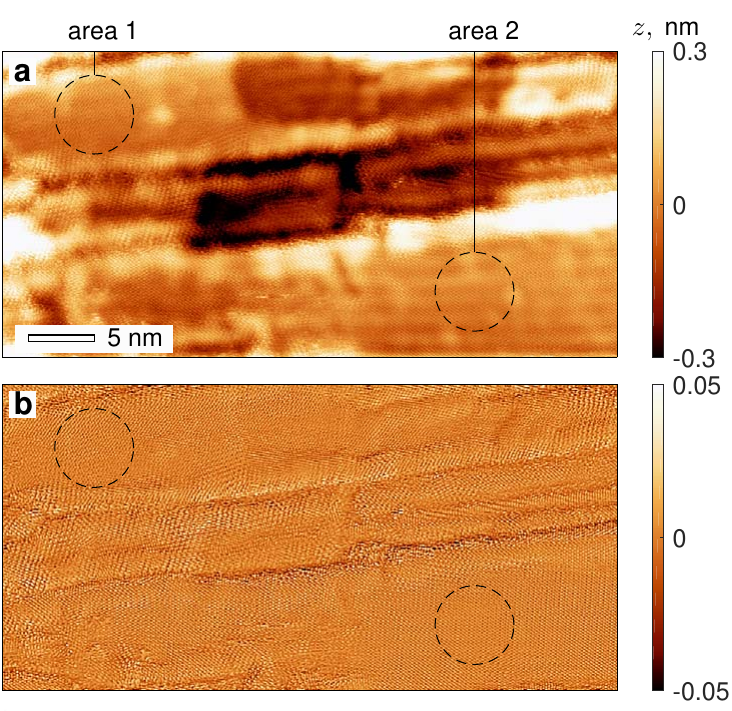}
\caption{\textbf{a} -- Raw topography image of a non-flat graphene film on the SiC/Si(001) surface (image size of $46.9\times 23.8$\,nm$^2$, $U=10\,$mV, $I=60\,$pA). \textbf{b} -- Map of the differential signal $D(x,y)$ acquired according to procedure~(\ref{Definition-DOG}) with the smoothing parameters are  $\sigma^{\,}_1=1\,$pxl, $\sigma^{\,}_2=2\,$pxl. The same image with higher resolution is shown in figure~S8 in the Supporting Information.
\label{Fig-09-Graphene}}
\end{figure}

\begin{figure}[h!]
\centering
\includegraphics[width=85 mm]{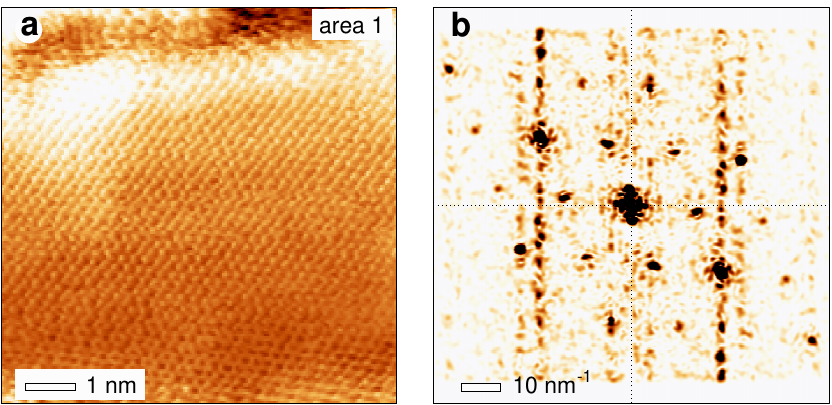}
\includegraphics[width=85 mm]{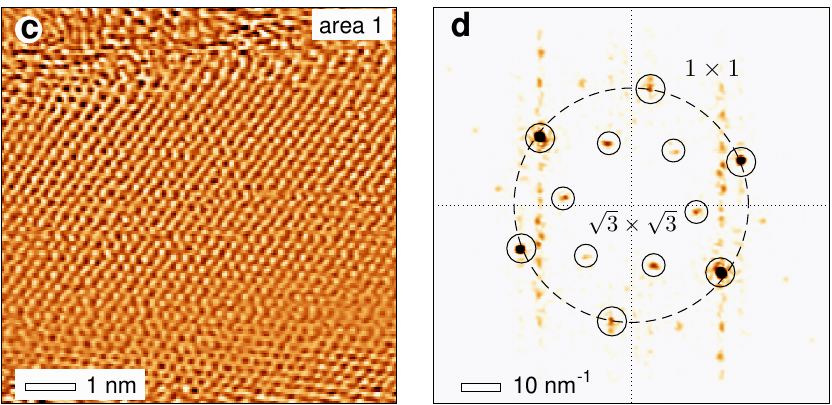}
\caption{\textbf{a}, \textbf{b} -- Raw topography image of the graphene layer within area 1 (panel a), and corresponding Fourier transform $|\hat{z}(k^{\,}_x,k^{\,}_y)|$ (panel b).
\textbf{c}, \textbf{d}  -- Map of the differential signal $D(x,y)$ (panel c) for the same area, and corresponding Fourier transform $|\hat{D}(k^{\,}_x,k^{\,}_y)|$ (panel d). From this point we use inverted color scheme when presenting results of the Fourier analysis: darker shades correspond to higher values and vice versa. The radius of the circle running through the first-order Fourier peaks is equal to $k^{\,}_0 = 4\pi /(\sqrt{3}\,a)\simeq 29.60\,$nm$^{-1}$, where $a=0.245\,$nm is the lattice constant for an ideal graphene film.
\label{Fig-10-Graphene1}}
\end{figure}

\begin{figure}[h!]
\centering
\includegraphics[width=85 mm]{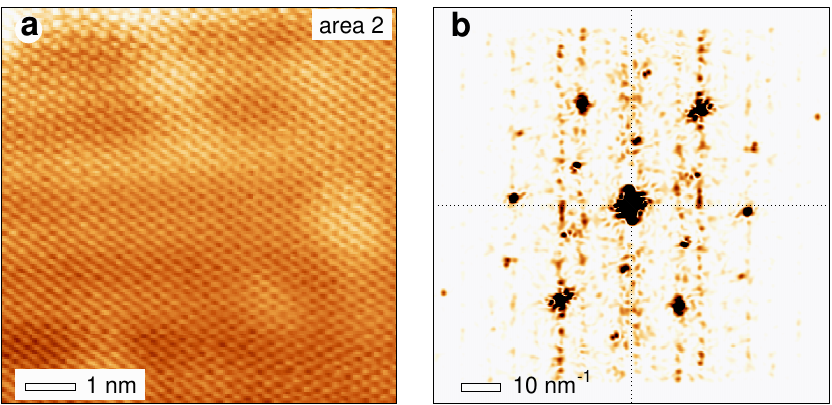}
\includegraphics[width=85 mm]{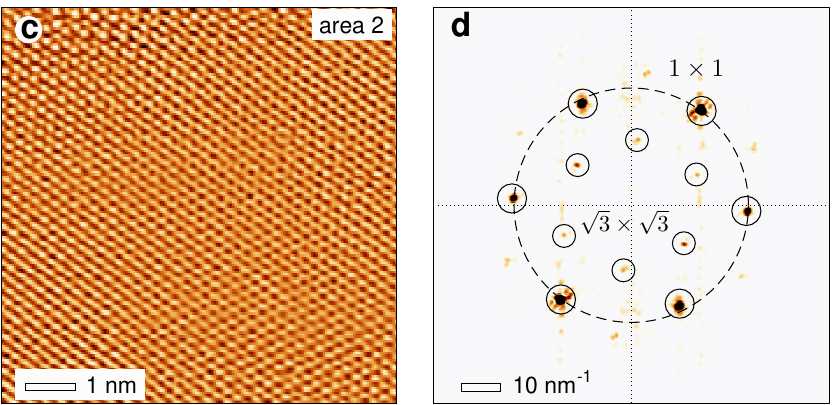}
\caption{\textbf{a}, \textbf{b} -- Raw topography image of the graphene layer within area 2 (panel a), and corresponding Fourier transform $|\hat{z}(k^{\,}_x,k^{\,}_y)|$ (panel b).
\textbf{c}, \textbf{d}  -- Map of the differential signal $D(x,y)$ (panel c) for the same area, and corresponding Fourier transform $|\hat{D}(k^{\,}_x,k^{\,}_y)|$ (panel d).
\label{Fig-11-Graphene2}}
\end{figure}

For deeper analysis, we consider two areas of interest (1 and 2) within the same antiphase domain (see figure~\ref{Fig-09-Graphene}a). A part of the raw topography image within the area 1 is shown in figure~\ref{Fig-10-Graphene1}a, and the same area after eliminating of surface rippling by procedure (\ref{Definition-DOG}) is illustrated in figure~\ref{Fig-10-Graphene1}c. The Fourier transforms corresponding to the raw and flattened images are presented in figures~\ref{Fig-10-Graphene1}b and \ref{Fig-10-Graphene1}d, respectively. One can easily recognize the Fourier peaks corresponding to the hexagonal graphene $1\times 1$ lattice and the peaks from the superstructure $(\sqrt{3}\times \sqrt{3})-R30^{\circ}$, related to the additional scattering of electronic waves on atomic defects and boundaries of nanodomains \cite{Mizes-1989,Tapaszto-2006,Tapaszto-2008}. Similar results for the area 2 are presented in figure~\ref{Fig-11-Graphene2}. It should be emphasized that the smoothing of raw images and the removal of non-flatness generally lead to the suppression of both the high-order Fourier peaks and the low-frequency noise without systematic broadening and/or shift of the first-order Fourier maxima.

\begin{figure}[h!]
\centering
\includegraphics[width=85 mm]{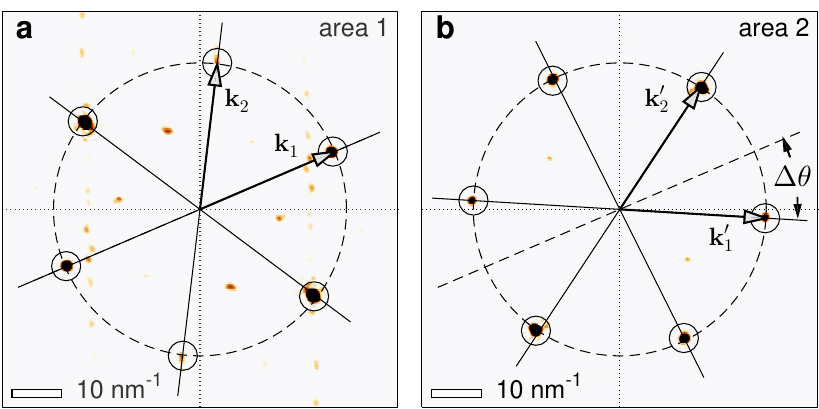}
\caption{\textbf{a}, \textbf{b}  -- Structure of the $k$-space for areas 1 (panel a) and 2 (panel b) and the first-order Fourier peaks after affine transformation (compare these images with figures~\ref{Fig-10-Graphene1}d and \ref{Fig-11-Graphene2}d). The basic vectors describing the hexagonal lattices within area 1 (${\bf k}^{\,}_1$ and ${\bf k}^{\,}_2$, panel a) and area 2 (${\bf k}^{\prime}_1$ and ${\bf k}^{\prime}_2$, panel b) are coherently rotated at the angle $\Delta\theta=26.6^{\circ}\pm 1.0^{\circ}$ relative to each other. The radius of the circle is equal to $k^{\,}_0 = 4\pi /(\sqrt{3}\,a)\simeq 29.60\,$nm$^{-1}$, where $a=0.245\,$nm is the lattice constant for an ideal graphene.
\label{Fig-12-Graphene3}}
\end{figure}

In order to minimize parasitic distortions caused by creep of piezo-scanner and thermal drift in the fast and slow scanning directions, we perform additional standard affine transformation for the maps of the differential signal within areas 1 and 2 using the same parameters. Figure~\ref{Fig-12-Graphene3} shows the structure of the $k$-space for the areas 1 and 2 after the affine transformation. The maps of the Fourier transforms for the differential signal allows us to determine precisely the angle between the translation vectors for areas 1 and 2: $\Delta\theta=26.6^{\circ}\pm 1.0^{\circ}$. This value is in agreement with the results of a recent paper \cite{Chaika-Nano-2015}: an asymmetric (with respect to the boundary between the neighboring nanoribbons) rotation of the translation vectors for local graphene lattices results in the formation of the periodic structure along the boundary with a twist angle of $26.57^{\circ}$. The rotation angle of $27^{\circ}$ was observed for different areas of the graphene layer grown on a polycrystalline Cu foil \cite{Huang-2011}. All these observations point to the fact that the formation of local structures at the graphene nanoribbon boundaries seems to be energetically favorable for the rotation angle, which differs from $30^{\circ}$. Such conclusions are also supported by the results of integral measurements for large-area (from $\mu$m to mm) graphene films on \mbox{$\beta$-SiC/Si(001)} by means of low-energy electron diffraction (LEED) and angle-resolved photoelectron spectroscopy (ARPES) \cite{Chaika-NanoRes-2013,Chaika-Nanotech-2014,Chaika-Nano-2019,Chaika-Nano-2015}. It is important to note that the precise determination of the orientation of lattices in different areas of the graphene layer becomes easier and more reliable if we apply the difference-of-Gaussians procedure to suppress low-$k$ noise and reduce the width of the first-order Fourier peaks (compare panels b and d in figures~\ref{Fig-10-Graphene1} and \ref{Fig-11-Graphene2}).

\begin{figure}[h!]
\begin{center}
\includegraphics[width=50 mm]{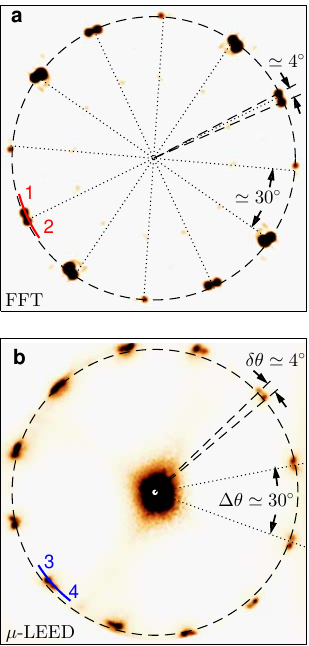}
\end{center}
\begin{center}
\includegraphics[width=82 mm]{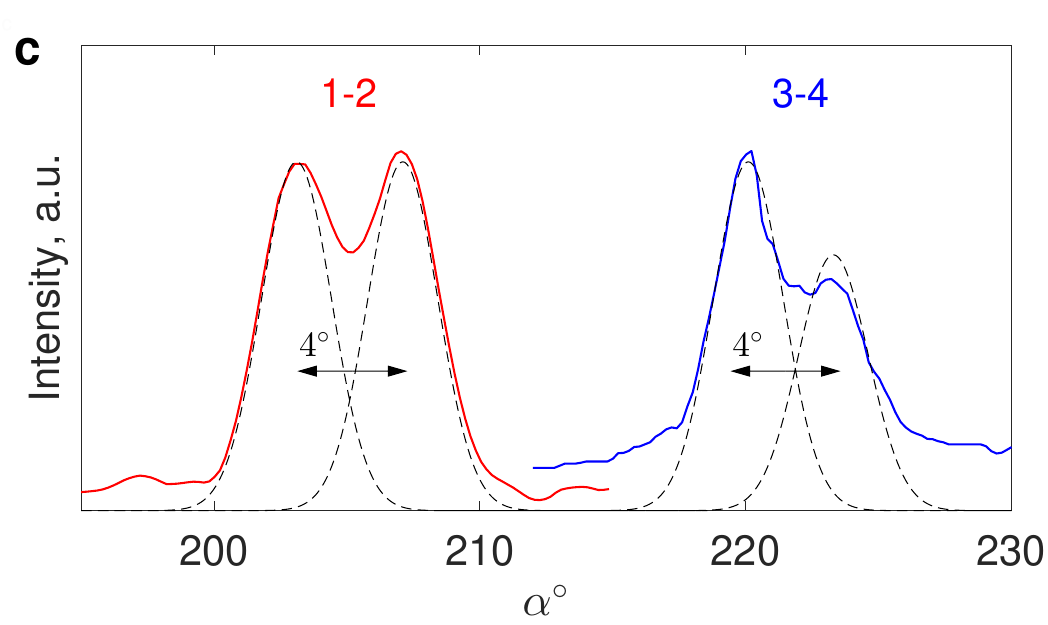}
\end{center}
\caption{\textbf{a} -- Combined map of the $k$-space, corresponding to the response of different areas in the nanostructured graphene layer, including rotations by $90^{\circ}$ due to the presence of antiphase domains in the \mbox{$\beta$-SiC/Si(001)} substrate.
\textbf{b} -- Raw micro-LEED pattern of nanostructured graphene on \mbox{$\beta$-SiC/Si(001)} surface, bright spots correspond to the first-order Fourier peaks from the $1\times 1$ lattice (beam energy $E=52\,$eV, sampling area 5~$\mu$m). Splitting of the first-order Fourier maxima due to the presence of the antiphase domains is well distinguishable (by the courtesy of A. Zakharov). \textbf{c}~-- Angular dependence of the intensities along lines 1-2 (panel a) and 3-4 (panel b) illustrating the splitting of the first-order Fourier maxima. Black dashed lines show schematically the decomposition of the splitted peaks into a combination of two localized Gaussian-like maxima.
\label{Fig-13-Graphene-LEED}}
\end{figure}

Figure~\ref{Fig-13-Graphene-LEED}a shows the expected structure of the $k$-space for a large-area graphene layer, containing nm-size areas with different orientations of the crystalline lattices (like areas 1 and 2 in figures~\ref{Fig-10-Graphene1}--\ref{Fig-12-Graphene3}) and $\mu$m-size antiphase domains. This image is a pixel-by-pixel summation of two local maps presented in figures \ref{Fig-12-Graphene3}a,b and the same maps rotated by $90^{\circ}$ in order to take into account two preferential orientations of the crystalline lattices in the substrate. The superposition of the Fourier reflexes apparently leads to (i) the appearance of a series of the first-order Fourier maxima positioned equidistantly at an angle of $30^{\circ}$; and (ii) the pronounced splitting of the first-order Fourier maxima. We estimate the mean angular width of each splitted spot: $\delta \theta\simeq 4^{\circ}$ (figure \ref{Fig-13-Graphene-LEED}c). This combined FFT image can be directly compared with the micro-LEED pattern for three-layered graphene on the \mbox{$\beta$-SiC/Si(001)} surface  (figure~\ref{Fig-13-Graphene-LEED}b). This map was recorded using a SPELEEM microscope (Elmitec GmbH) (see details in \cite{Chaika-NanoRes-2013}). The Fourier peaks splitting is also visible in standard LEED and ARPES experiments with a sampling area on the order of few mm \cite{Chaika-NanoRes-2013,Chaika-Nano-2019,Chaika-Nano-2015}.  Such fantastic agreement between the results of local measurements (figure~\ref{Fig-13-Graphene-LEED}a) and integral measurements (figure~\ref{Fig-13-Graphene-LEED}b) points to the exceptional importance of numerical methods of analysis of high-resolution STM data (in particular, the difference-of-Gaussians procedure) for consistent processing of experimental data acquired by various techniques.

\section{Conclusion}

We have shown that simple and effective difference-of-Gaussians approach can be applied for convenient presentation and deeper analysis of high-resolution data obtained by scanning probe microscopy in non-standard cases. This algorithm is quite stable and allows one to minimize both global slope and surface curvature for studied samples and obtain flattened images for stepped, facetted, and non-flat surfaces for further analysis of surface corrugations.  In contrast to the results of the first-order numerical differentiation ($\partial z/\partial x$ and $\partial z/\partial y$), the difference-of-Gaussians approach returns the output pattern, which is close to the Laplacian ($\partial^2/\partial x^2 + \partial^2/\partial y^2$) of the blurred topography image. We have argued that the difference-of-Gaussians approach can be especially effective for 'on-the-fly' express analysis of SPM data and/or detailed studies, which require high-precision measurements of metrological quality. In particular, we have demonstrated that this approach makes it possible to (i) visualize crystalline lattices for all faces of the facetted surface or 3D crystallites simultaneously; (ii) visualize crystalline lattices on the non-flat surfaces of graphene and graphene-like structures, including the regions of highly corrugated nanoribbon boundaries, uncover details of their atomic structure and determine the relative rotation of the local lattices in different areas with high precision (better than $1^{\circ}$). We have shown that the maps of the differential signal obtained by the difference-of-Gaussians approach provide qualitatively the same atomic features as the projection of the topography image onto the $x-y$ scanning plane. The difference-of-Gaussians approach can be applied to the numerical analysis of images acquired by other experimental techniques such as scanning/transmission electron microscopy, angle-resolved photoelectron spectroscopy, etc.

\section*{Acknowlednements}

The authors thank Ilya Shereshevsky, Vasily Stolyarov and Goran Karapetrov for valuable comments, Alexei Zakharov for providing us with micro-LEED data, and Daniil Bratashov for his consultancy concerning the Gwyddion software. This work was financially supported by Russian State Contracts of Institute for Physics of Microstructures RAS (FFUF-2024-0020, preparation of samples and measurements) and Osipyan Institute of Solid State Physics RAS (preparation of samples and measurements), and by the Grant from the Ministry of Science and Higher Education of the Russian Federation No. 075-15-2024-632 (programming and data analysis).

\newpage

\setcounter{page}{1}
\setcounter{figure}{0}

\renewcommand{\thefigure}{S\arabic{figure}}
\renewcommand{\thepage}{S\arabic{page}}
\renewcommand{\theequation}{S\arabic{equation}}

\begin{center}
{\large {\bf \textcolor[rgb]{0.00,0.00,0.55}{Supporting Information}}}
\end{center}

\begin{center}
{\large {\bf Visualization of Atomic Structures on Faceted and Non-Flat Surfaces by Difference-of-Gaussians Approach}}
\end{center}

\begin{center}
A. Yu. Aladyshkin, A. N. Chaika, V. N. Semenov, A. S. Aladyshkina, S. I. Bozhko,  A. M. Ionov
\end{center}

\begin{figure*}[t!]
\centering
\includegraphics[width=120 mm]{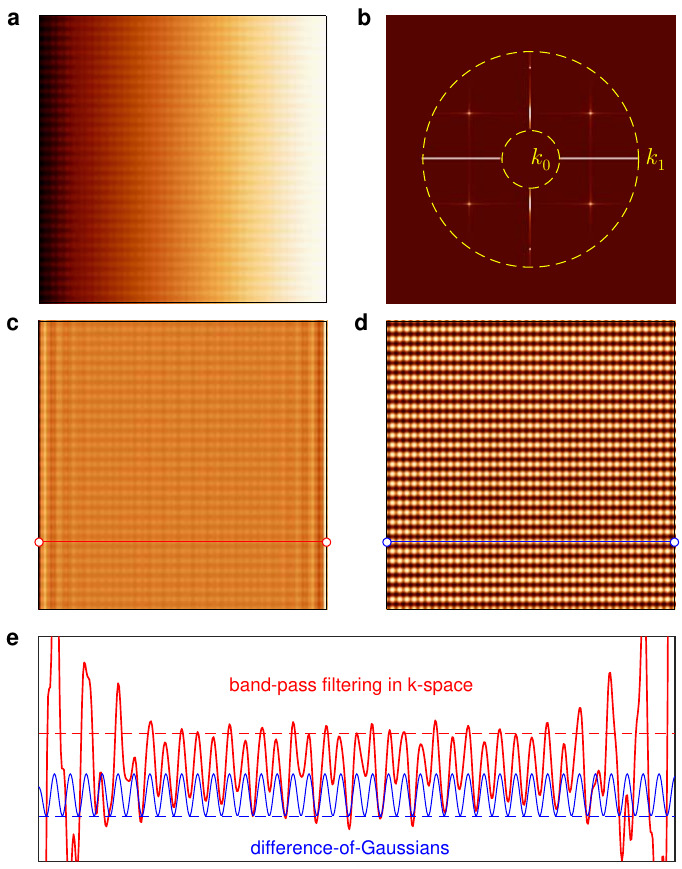}
\caption{\textbf{a}~-- Model topography image which mimics a tilted face of a crystal with rectangular atomic lattice ($40 \times 30$ unit cells,  $600 \times 600$ pixels, no noise). \textbf{b}~-- Structure of $k$-space, corresponding to the image in panel a after band-pass filtering.  \textbf{c}~-- Real part of restored topography image after band-pass piece-wise filtering in $k$-space (only the components with $k$ values obeying the criterion $k^{\,}_0<|k|<k^{\,}_1$ are taken into account). \textbf{d}~-- Map of the differential signal $D(x,y)$ acquired by means of the difference-of-gaussians procedure (see Eq.~(9) in the main paper), smoothing parameters are $\sigma^{\,}_1=1\,$pxl and $\sigma^{\,}_2=4 \sigma^{\,}_1$. \textbf{e}~-- Cross-sectional views of the images in panel c and d along red and blue horizontal lines, respectively. It is easy to conclude that the procedure of the band-pass piece-wise filtering in $k$-space results in appearance of visible variations in the amplitude of periodic signal unlike for the difference-of-gaussians approach.
\label{Fig-S07a}}
\end{figure*}

\begin{figure*}[t!]
\centering
\includegraphics[width=120 mm]{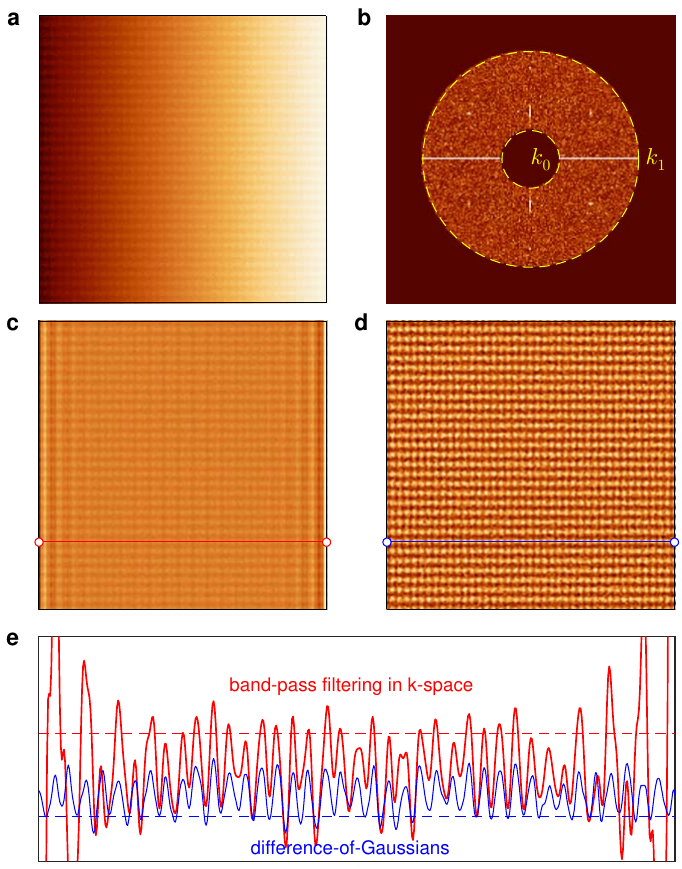}
\caption{The same as figure S1, but for noisy model profile.
\label{Fig-S07b}}
\end{figure*}

\bigskip
Figures \ref{Fig-S07a} and \ref{Fig-S07b} illustrate the effectiveness of two different procedures (difference-of-gaussians in real space and Fourier filtration in $k$-space by means of a standard rectangular band-pass filter) for elimination of global slope and visualization of atomic structure for model tilted surface with rectangular lattice.

\begin{figure*}[h!]
\centering
\includegraphics[width=100 mm]{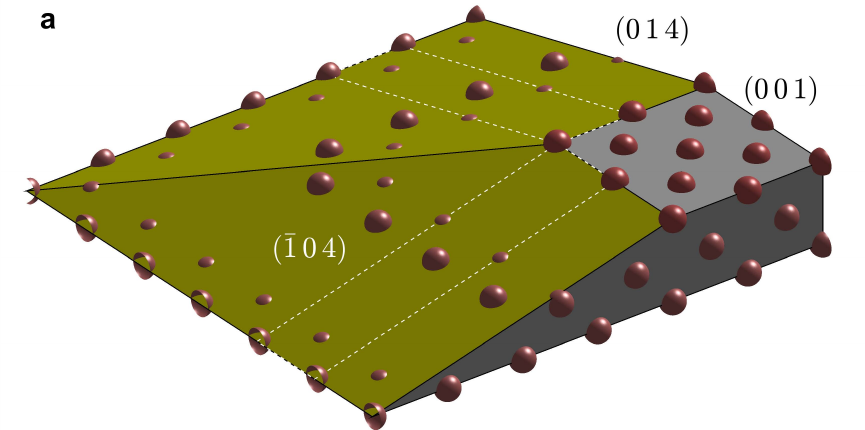}
\includegraphics[width=90 mm]{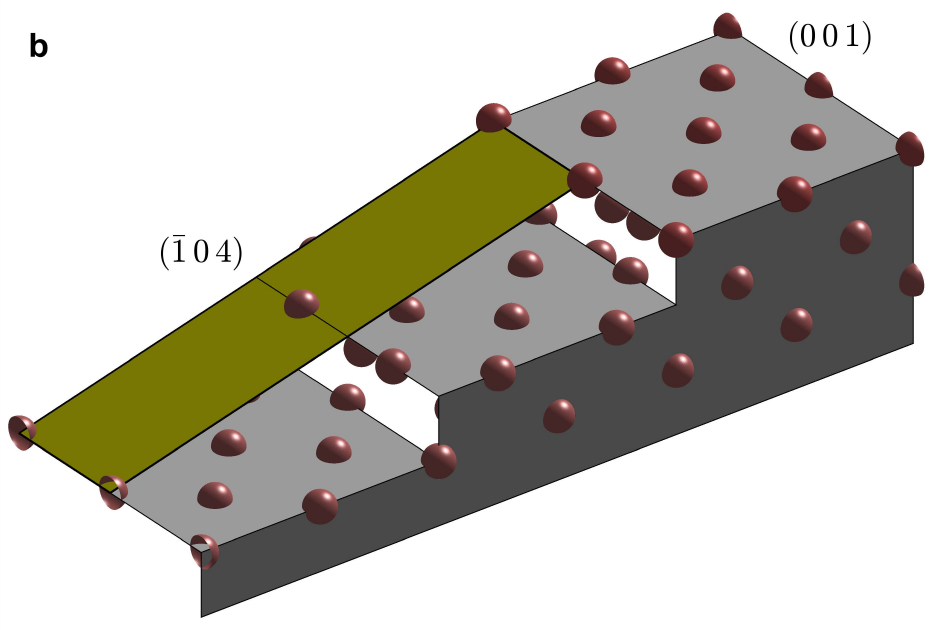}
\caption{\textbf{a, b}~-- Model presentation of the different faces for ideal Cu single crystal (fcc structure, lattice constant 0.361\,nm). Dashed rectangles in panel a depict unit cells for the $c(4\times 1)$ reconstruction appearing at $\{0\,1\,4\}$ faces after oxidation.  Oxygen atoms are expected to be positioned at (0\,0\,1) planes between Cu atoms (not shown for clarity). Overlayer model as well as top and side views for reconstructed Cu(0\,1\,4)--O surface is presented in Chaika \emph{et al.} Surface Science, v. 602, 2078-2088 (2008).
\label{Fig-S01}}
\end{figure*}

\medskip
Figure \ref{Fig-S01} shows three-dimensional model of ideal Cu crystal and mutual arrangements of different faces: $(0\,0\,1)$, $(\bar{1}\,0\,4)$ and $(0\,1\,4)$.

\begin{figure*}[h!]
\centering
\includegraphics[width=150 mm]{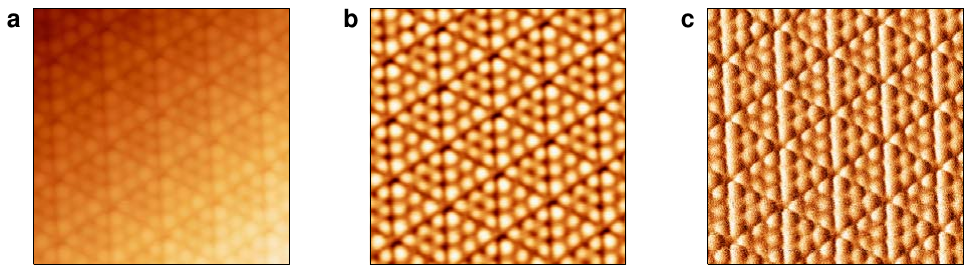}
\caption{\textbf{a}~-- Raw topography image of tilted reconstructed Si(1\,1\,1)$7\times 7$ surface (\mbox{$11.4\times 11.4\,$nm$^2$}, \mbox{$U=-0.90\,$V}, \mbox{$I=200\,$pA}). \textbf{b}~-- Map of the differential signal $D(x,y)$ acquired by means of the difference-of-gaussians procedure (see Eq.~(9) in the main paper), smoothing parameters are $\sigma^{\,}_1=0.05\,$nm and $\sigma^{\,}_2=0.25\,$nm. \textbf{c}~-- Map of the differential signal $\partial z(x,y)/\partial x$ acquired by numerical differentiation along the fast-scanning direction. Image in panel c is similar to the map of variations of the tunneling current $I(x,y)$, which characterizes non-ideal operation of feedback loop.
\label{Fig-S02}}
\end{figure*}

\begin{figure*}[h!]
\centering
\includegraphics[width=150 mm]{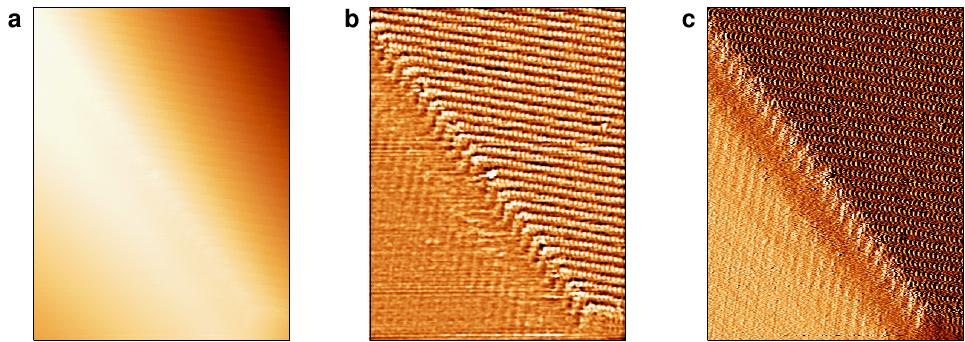}
\caption{\textbf{a}~-- Raw topography image of tilted reconstructed Cu(1\,1\,5)--O surface (\mbox{$19.4\times 29.4\,$nm$^2$}, \mbox{$U=-10\,$mV}, \mbox{$I=150\,$pA}), see figure~4 in the main paper. \textbf{b}~-- Map of the differential signal $D(x,y)$ acquired by means of the difference-of-gaussians procedure (see Eq.~(9) in the main paper), smoothing parameters are $\sigma^{\,}_1=0.05\,$nm and $\sigma^{\,}_2=0.22\,$nm. \textbf{c}~-- Map of the differential signal $\partial z(x,y)/\partial x$ acquired by numerical differentiation along the fast-scanning direction.
\label{Fig-S03}}
\end{figure*}

\begin{figure*}[h!]
\centering
\includegraphics[width=130 mm]{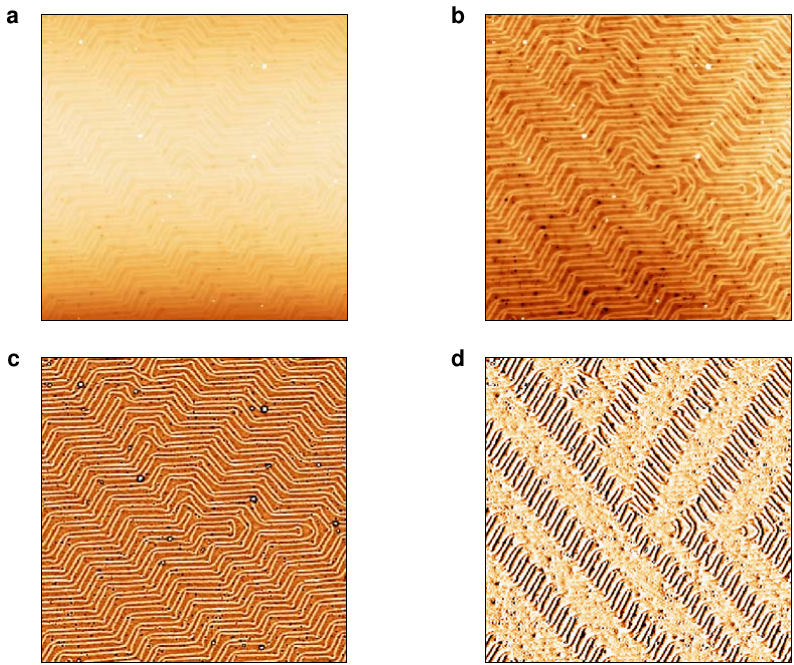}
\caption{\textbf{a} -- Raw topography image of Au(1\,1\,1) film with remarkable herring-bone $22\times \sqrt{3}$ reconstruction ($150\times 150$\,nm$^2$, $U=0.50\,$V, $I=20\,$pA). \textbf{b} -- Aligned topography image prepared by linewise correction, which minimizes non-linear background, for the image in the panel a. \textbf{c}~-- Map of the differential signal $D(x,y)$ acquired by means of the difference-of-gaussians procedure (see Eq.~(9) in the main paper) for the image in the panel a, smoothing parameters are $\sigma^{\,}_1=0.2\,$nm and $\sigma^{\,}_2=0.4\,$nm. \textbf{d}~-- Map of the differential signal $\partial z(x,y)/\partial x$ acquired by numerical differentiation along the fast-scanning direction for the image in the panel a.
\label{Fig-S04}}
\end{figure*}

\begin{figure*}[h!]
\centering
\includegraphics[width=160 mm]{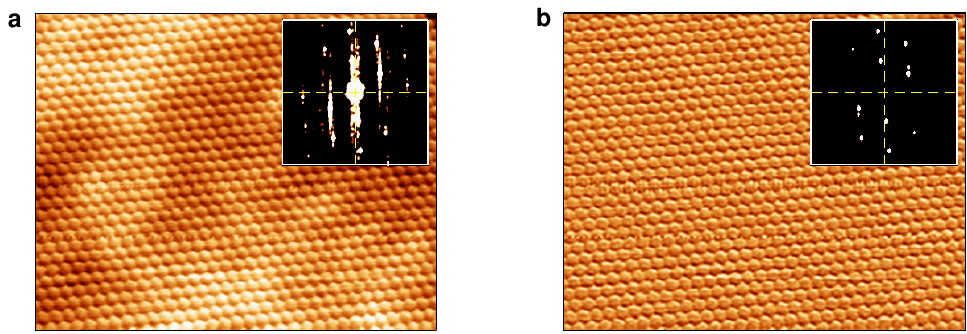}
\caption{\textbf{a} -- Raw topography image of non-flat thin Pb(1\,1\,1) film ($11.1\times 8.8$\,nm$^2$, $U=-10\,$mV, $I=40\,$pA). \textbf{b}~-- Map of the differential signal $D(x,y)$ acquired by means of the difference-of-gaussians procedure (see Eq.~(9) in the main paper), smoothing parameters are $\sigma^{\,}_1=0.02\,$nm and $\sigma^{\,}_2=0.04\,$nm. Insets show the Fourier transforms for the corresponding images (image size $50\times 50$\,nm$^{-2}$).
\label{Fig-S05}}
\end{figure*}

\medskip
Figures \ref{Fig-S02}--\ref{Fig-S05} illustrates the effectiveness of two different procedures (difference-of-gaussians and direct numerical differentiating along $x$-axis) for tilt elimination and visualization of atomic structures for the following cases:

\smallskip
-- reconstruction Si(1\,1\,1)$7\times 7$ (figure \ref{Fig-S02});

\smallskip
-- reconstruction Cu(1\,0\,4)--O (figure \ref{Fig-S03});

\smallskip
-- herring-bone reconstruction Au(1\,1\,1)$22\times \sqrt{3}$ (figure \ref{Fig-S04});

\smallskip
-- hexagonal lattice on top of Pb(1\,1\,1) film (figure \ref{Fig-S05}).

\begin{figure*}[h!]
\centering
\includegraphics[width=160 mm]{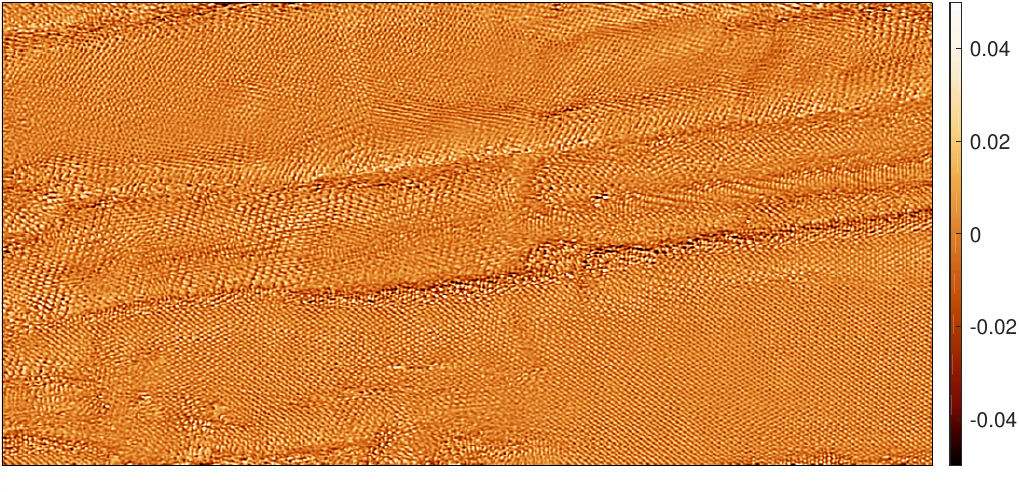}
\caption{Map of the differential signal $D(x,y)$ for the nanostructured graphene film on SiC/Si(001) surface (the same image with lower resolution is shown in figure~9b of the main paper).
\label{Fig-S06}}
\end{figure*}

\medskip
Figure \ref{Fig-S06} reproduces figure \ref{Fig-09-Graphene}b from the main paper with better resolution.

\newpage

\clearpage

\end{document}